\documentclass[journal,onecolumn,draftclsnofoot,english,12pt]{IEEEtran}
\usepackage[T1]{fontenc}
\usepackage[justification=centering]{caption}
\usepackage{lettrine}
\usepackage{float}
\usepackage{amsmath}
\allowdisplaybreaks[4]
\usepackage{babel}
\usepackage{amssymb}
\usepackage{graphicx}
\usepackage{epstopdf}
\usepackage{array}
\usepackage{cancel}
\usepackage{float}
\usepackage{amsthm}
\usepackage{amsmath}
\usepackage{array}
\usepackage{amssymb}
\usepackage{stackrel}
\usepackage{multirow}
\usepackage{colortbl}
\usepackage{tabu}
\usepackage{setspace}
\usepackage{url}
\usepackage{algorithm}
\usepackage{algorithmic}
\usepackage{booktabs}
\usepackage{cite}
\usepackage{color}

\usepackage{subfigure}
\usepackage[numbers,sort&compress]{natbib}
\newcolumntype{Y}{>{\centering\arraybackslash}X}

\makeatletter

\floatstyle{ruled}
\newfloat{algorithm}{tbp}{loa}
\providecommand{\algorithmname}{Algorithm}
\floatname{algorithm}{\protect\algorithmname}

\theoremstyle{plain}

\theoremstyle{definition}

\theoremstyle{plain}

\theoremstyle{plain}

\IEEEoverridecommandlockouts
\usepackage{ifpdf}
\usepackage{cite}
\usepackage{amsmath}
\@ifundefined{showcaptionsetup}{}{%
\PassOptionsToPackage{caption=false}{subfig}}
\makeatother

\usepackage{babel}

\begin{document}
\title{Computation and Privacy Protection for Satellite-Ground Digital Twin Networks}
\author{Yongkang Gong,~\emph{Student Member, IEEE}, Haipeng Yao,~\emph{Senior Member, IEEE},
Xiaonan Liu,~\emph{Member, IEEE},
Mehdi Bennis,~\emph{Fellow, IEEE},
Arumugam Nallanathan,~\emph{Fellow, IEEE},
and Zhu Han,~\emph{Fellow,~IEEE}

\thanks{Y. Gong and H. Yao are with the School of Information and Communication Engineering, BUPT, Beijing, 100876, China (e-mail: \{yongkanggong, yaohaipeng\}@bupt.edu.cn).\par
X. Liu and A. Nallanathan are with the Communication Systems Research group in Queen Mary University of London, QMUL, London, E1 4NS, the United Kingdom (e-mail: \{x.l.liu, a.nallanathan\}@qmul.ac.uk).\par
M. Bennis is with the Center for Wireless Communications, University of Oulu, 90014 Oulu, Finland (e-mail: mehdi.bennis@oulu.fi).\par
Z. Han is with the School of Electrical and Computer Engineering, University of Houston, Houston, TX 77004 USA (e-mail: zhan2@uh.edu).
}}

\maketitle

\begin{abstract}
Satellite-ground integrated digital twin networks (SGIDTNs) are regarded as innovative network architectures for reducing network congestion, enabling nearly-instant data mapping from the physical world to digital systems, and offering ubiquitous intelligence services to terrestrial users. However, the challenges, such as the pricing policy, the stochastic task arrivals, the time-varying satellite locations, mutual channel interference, and resource scheduling mechanisms between the users and cloud servers, are critical for improving quality of service in SGIDTNs. Hence, we establish a blockchain-aided Stackelberg game model for maximizing the pricing profits and network throughput in terms of minimizing overhead of privacy protection, thus performing computation offloading, decreasing channel interference, and improving privacy protection. Next, we propose a Lyapunov stability theory-based model-agnostic meta-learning aided multi-agent deep federated reinforcement learning (MAML-MADFRL) framework for optimizing the CPU cycle frequency, channel selection, task-offloading decision, block size, and cloud server price, which facilitate the integration of communication, computation, and block resources. Subsequently, the extensive performance analyses show that the proposed MAML-MADFRL algorithm can strengthen the privacy protection via the transaction verification mechanism, approach the optimal time average penalty, and fulfill the long-term average queue size via lower computational complexity. Finally, our simulation results indicate that the proposed MAML-MADFRL learning framework is superior to the existing baseline methods in terms of network throughput, channel interference, cloud server profits, and privacy overhead.
\end{abstract}


\IEEEpeerreviewmaketitle

\section{Introduction}
\lettrine[lines=2]{A}lthough terrestrial networks \cite{feng2017hetnet} can support high data rate and large-scale terminal devices access, it is hard to provide seamless, global, and uniform coverage for low user-density regions and to fulfill the rapid proliferation of mobile applications in the sixth-generation wireless communication systems (6G) \cite{gong2022decentralized}. Fortunately, satellite-ground integrated networks (SGINs) \cite{gong2022computation} can fill the coverage-holes to support global coverage, especially in remote suburbs, oceans, and deserts. Specifically, satellite networks (e.g., Starlink) consist of multiple low earth orbits (LEO), medium earth orbits, and geostationary earth orbits \cite{song2021aerial}, which provide global coverage, moderate relay transmission, and task processing functionalities for terrestrial users \cite{gong2022computation1}.

There have been some related contributions about edge task offloading. Specifically, Luo \textit{et al}. \cite{luo2022cost} proposed edge server network design algorithms to balance the construction cost and the network density for edge networks. Next, Alnoman \textit{et al}. \cite{alnoman2019computing} explored a sharing and disjoint cloud-edge system to minimize the response time via dynamic programming and exhaustive searching methods. Fan \textit{et al}. \cite{fan2021game} established a game-theory based multi-type computation offloading mechanism to balance the task computing delay among multiple base stations (BSs).

However, when terrestrial users offload tasks to macro base station (MBS) servers, it may cause severe channel interference among different edge networks, bring huge transmission energy consumption and reduce the ability of processing number of tasks \cite{wang2017taking}. Moreover, the above works do not consider the corresponding cloud servers' pricing mechanisms and the processed number of task bits. Fortunately, digital twin (DT) \cite{mihai2022digital} is regarded as a novel technique, which can enable instant wireless connectivity as well as data mapping services, and shorten the gap among physical utilities and digital systems \cite{khan2022digital}. Furthermore, Lu \textit{et al}. \cite{lu2021adaptive} formulated the edge association problem including DT placement and DT migration, and then employed the deep reinforcement learning (DRL) and transfer learning mechanisms to improve the convergent rate. Huynh \textit{et al.} \cite{van2022edge} established a multi-access edge computing-based ultra-reliable and low latency communications architecture to optimize the offloading portions, bandwidth and server computation capability, which can improve latency and reliability in metaverse applications.

However, the privacy protection for task offloading and lack of mutual trust among terrestrial users impede resource sharing and cooperation \cite{hou2022environment}. To solve these challenges, blockchain technologies can be widely deployed to achieve transaction verification and privacy protection functionalities among substantial users. Qiu \textit{et al.} \cite{qiu2020networking} improved the proof of work via proposing a blockchain-assisted collective Q-learning method. Furthermore, Cao \textit{et al.} \cite{cao2022hierarchical} conceived a blockchain-aided software-defined energy network and designed a distributed energy smart contract to guarantee transactions reliably and accurately.

Despite the advantage of blockchain technique \cite{li2023cryptoeconomics} for improving privacy protection, the learning efficiency needs to be further explored for dynamic network environment. Nguyen \textit{et al.} \cite{hieu2022deep} utilized DRL to minimize the latency and mining cost of machine learning model owner. Furthermore, Du \textit{et al.} \cite{du2021resource} utilized an asynchronous advantage actor-critic algorithm to obtain the optimal resource pricing and allocation, and used the prospect theory to balance risks and rewards. Ma \textit{et al.} \cite{ma2022blockchain} established an autonomous control platform to optimize network resources, adjust power services, and maximize the profits for consumers and operators. However, these mentioned methods cause large transmission overhead as well as low learning efficacy and may leak user privacy. Hence, a model-agnostic meta-learning (MAML) framework is introduced to quickly adapt to new tasks from small samples, which can greatly improve the learning efficiency and accelerate the convergent rate. Furthermore, deep federated reinforcement learning (DFRL) is deployed to execute the task scheduling and strengthen the privacy protection for dynamic network environment.

Inspired by above challenges, we conceive satellite-ground integrated digital twin networks (SGIDTNs) scenario for computation offloading, alleviating channel interference, and improving privacy protection among users under dynamic network environment, and then maximize the cloud servers profits for time-varying DT computation capability. The main contributions are presented as follows.
\begin{itemize}
\item We envision a blockchain-aided two-stage Stackelberg game model to maximize the processed number of task bits and cloud servers profits. Integrated with in-orbit intelligent computation, it helps the network adapt to the stochastic task arrivals, the time-varying LEO locations, the cloud server price, and the DT computation frequency. Furthermore, it decouples the variable coupling for the long-term task queue and the short-term task offloading.
\item We propose a Lyapunov stability theory-based MAML-MADFRL framework to optimize DT computation frequency, allocate wireless channel, execute task offloading, choose block size, and obtain the optimal price for cloud servers. Specifically, the Lyapunov-based policy is convoked to decouple the long-term task queues. Next, the proposed MAML-MADFRL framework is utilized to process the computation offloading, channel interference, and privacy protection. Moreover, the MAML-MADFRL framework can obtain the optimal price for cloud servers in the second-stage Stackelberg game.
\item Massive theoretical analyses show that the proposed Lyapunov-based MAML-MADFRL framework can validate the transaction process, approach the optimal performance, and satisfy the long-term task queue constraint via lower computation complexity. Furthermore, extensive simulation results indicate that the proposed learning framework is superior to the traditional baselines, such as multi-agent random task offloading (MARTO), multi-agent mean CPU cycle frequency (MAMCC) and multi-agent greedy channel selection (MAGCS).
\end{itemize}

The structure of the paper is concluded as follows. We list some related works in Section II. Moreover, Section III establishes the blockchain-aided system model in the SGIDTNs scenario. Next, we introduce the corresponding Lyapunov stability theory-based MAML-MADFRL learning framework in Section IV. Furthermore, the transaction verification, the computational complexity of the algorithm, the convergent rate, and the task queue constraints are demonstrated in Section V. Subsequently, massive simulation results are presented in Section VI. Finally, we conclude this paper in Section VII.

\section{Related Works}
\ \ Blockchain-aided SGIDTNs are considered as the prospective network architecture to achieve flexible deployment, global coverage, and cognitive capability. Specifically, Cao \textit{et al.} \cite{cao2021hap} conceived a transmission control policy for ground-air-space and ground-to-space links and maximized the overall network throughput. Guo \textit{et al.} \cite{guo2021survey} provided a detailed survey about network security on space-air-ground-sea network. Fan \textit{et al.} \cite{fan2022network} processed the network selection via evolutionary game and utilized the deep deterministic policy gradient to handle high-dimensional action spaces. However, the aforementioned works do not consider the real-time task processing between physical unities and digital systems.

Recently, DT can be utilized to accelerate the wireless network evolution and map the task data to digital systems. Lu \textit{et al.} \cite{lu2020low} introduced DT to wireless networks and proposed a learning framework to balance the learning efficiency and time cost. Bellavista \textit{et al.} \cite{bellavista2021application} processed the application-enabled DT equipment and applied software-defined networking to explore communication mechanisms. Lei \textit{et al.} \cite{lei2021toward} established a DT-based thermal power plant and explored the web-based architecture and control algorithm.

However, the lack of mutual trust as well as privacy protection, and low training efficiency reduce the quality of service (QoS). Hence, some research works focused on blockchain and federated learning \cite{ma2022federated}. Specifically, Qu \textit{et al.} \cite{qu2020blockchained} developed a decentralized cognitive computing paradigm and utilized the blockchain-aided federated learning technique to solve data island and incentive mechanism. Cui \textit{et al.} \cite{cui2022fast} proposed a blockchain-aided compressed federated learning framework to maximize the final model accuracy and minimize the training loss. Wang \textit{et al.} \cite{wang2019edge} designed an "In-Edge AI" framework to reduce system communication overhead and integrated DRL with FL to optimize communication, computation as well as caching resources \cite{zhang2021deep}. Nevertheless, aforementioned works cannot guarantee fast model adaptability from small batches of samples.

In contrast with aforementioned works, we propose a Lyapunov stability theory-based MAML-MADFRL learning framework to decouple the long-term task queues and accelerate the learning rate for multiple task scenarios in SGIDTNs, which can adapt to dynamic network environment and maximize the network throughput as well as cloud server profits in two-stage Stackelberg game process.

\section{System Model}
\subsection{SGIDTNs Blockchain Scenario}
\begin{figure}[!t]
\centering{}
\setlength{\abovecaptionskip}{0.5cm}
\includegraphics[width=12cm, height=8cm]{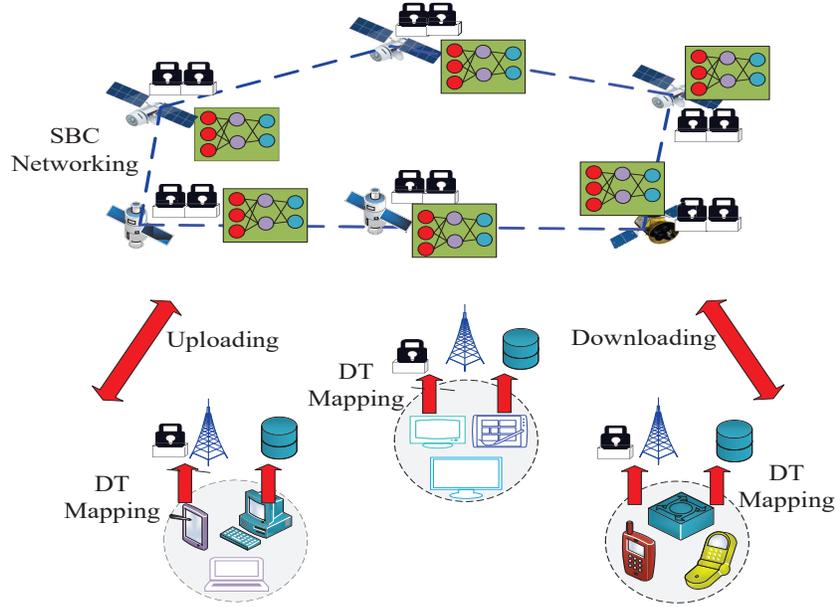}
\caption{The SGIDTNs blockchain scenario.}
\setlength{\belowdisplayskip}{-0.1cm}
\label{The SAG-integrated three-layer network}
\end{figure}

\ \ As shown in Fig. \ref{The SAG-integrated three-layer network}, we introduce the SGIDTNs blockchain scenario, which is deployed to achieve computation offloading and privacy protection for multiple ground devices (GDs). The network scenario consists of two main layers, i.e., satellite networks and terrestrial networks. Specifically, the terrestrial network is composed of multiple MBSs, the sets of which are represented as $\mathbb{N}=\{1,2,...,n,...,N\}$. Moreover, each $n$ overlays $\mathbb{M}=\{1,2,...,m,...,M\}$ GDs and all $M$ GDs execute instant wireless access and reliable data mapping from physical entities to digital space via the DT technology.

Subsequently, the satellite networks consist of massive LEO satellites, whose sets are denoted as $O=\{1,2,...,o,...,O\}$. Additionally, all LEO satellites can provide global communication coverage and seamless connectivity for GDs. Meanwhile, tasks are offloaded to LEO to relieve computation pressure while protecting data privacy via federated aggregation and issuing mechanisms. Specifically, as multiple DTs lack mutual trust and cannot share local data, we propose a blockchain-aided federated aggregation policy to execute model parameters aggregation and issue network parameters, which can not only achieve computation offloading and improve training efficiency, but also further protect data privacy.

Next, when multiple DTs offload their tasks to LEO satellites, cloud providers can set own price to earn more economic profits, and then massive followers can determine own service demand after obtaining the price strategy of cloud providers. Hence, the resource orchestration and pricing strategy between cloud providers and multiple DTs can be regarded as a Stackelberg game process, and we need to find the optimal policy to fulfill the service requirements for both cloud providers and DTs.

Meanwhile, blockchain is a novel distributed ledger technology, which can be utilized to strengthen mutual trust among massive GDs. Specifically, when tasks are offloaded to LEO servers, the corresponding LEO server can help verify model parameters and issue them to corresponding ground DTs, which guarantee secure network transaction.

\subsection{DT Computation Model}

\ \ Furthermore, we can divide the Stackelberg game process into two stages, i.e., the follower stage and the provider stage. First, massive DTs tend to maximize the number of processed bits in terms of minimum blockchain verification overhead, and detailed computation offloading and blockchain verification processes are shown as follows.

For each time slot $t$, the DT $m$ in the ${{n}^{th}}$ MBS receives one task $A_{n,m}^{t}$ and assuming that the second moment is limited, i.e., $\mathbb{E}\left([A_{n,m}^{t}]^{2}\right)={{\eta}_{n,m}}<\infty$. Moreover, ${{\eta }_{n,m}}$ can be obtained via collecting previous network statistics. Next, when each DT processes the task locally, the processed number of task bits is calculated as
\begin{align}
D_{n,m}^{t1}=\frac{f_{n,m}^{t}}{w}T,
\end{align}
where $f_{n,m}^{t}$ is the CPU cycle frequency of each DT, $w$ is the allocated number of CPU cycles while processing one bit task, and $T$ is time slot duration.
\subsection{Computation Offloading Model}
\ \ When DTs offload their tasks to LEO satellites, stochastic task arrivals, dynamic LEO locations, and mutual channel interference among different ground edge networks cause severe impact on computation offloading and privacy protection. Moreover, the loss of the communication link between DT $m$ and LEO $o$ for the ${{n}^{th}}$ MBS is denoted as
\begin{equation}
\begin{aligned}
& L_{n,m}^{t}=20\log \left( {}^{4\pi {{f}_{c}}\sqrt{x_{n,m,o}^{2}+y_{n,m,o}^{2}}}/{}_{c} \right)+p_{n,m,o}^{LoS}\varepsilon _{n,m,o}^{LoS} \\
&\ \ \ \ \ \ +\left( 1-p_{n,m,o}^{LoS} \right)\varepsilon _{n,m,o}^{NLoS},
\end{aligned}
\end{equation}
where ${{\text{f}}_{c}}$ is the carrier frequency and $c$ is the speed of light, ${{x}_{n,m,o}}$ is the horizontal distance between DT $m$ and LEO $o$, ${{y}_{n,m,o}}$ is the flight altitude for each LEO satellite $o$. Additionally, $\varepsilon_{n,m,o}^{LoS}$ and $\varepsilon_{n,m,o}^{NLoS}$ are related additional path loss imposed on free space propagation from line-of-sight (LoS) and non-line-of-sight (NLoS), respectively. Furthermore, the corresponding LoS propagation probability is denoted as
\begin{align}
p_{n,m,o}^{LoS}={}^{1}/{}_{1+b1\exp \left\{ -b2[arctac\left( \frac{y_{n,m,o}^{{}}}{{{x}_{n,m,o}}} \right)-b1] \right\}},
\end{align}
where $b1$ and $b2$ are corresponding constants \cite{gong2022computation}, which are obtained via interacting with dynamic environment. Subsequently, when multiple DTs offload their tasks to LEO satellites, it can cause massive mutual interference among different edge networks. We assume that multiple DTs transmit their tasks via orthogonal frequency division multiple access (OFDMA), which causes interference among multiple edge networks. Hence, for the ${{m}^{th}}$ DT in the ${{n}^{th}}$ MBS, the channel interference is represented as
\begin{align}
{{I}_{n,m,r}}=\sum\limits_{q=1,q\ne n}^{N}{\sum\limits_{m=1}^{M}{{{\beta }_{q,m,r}}{{P}_{q,m,r}}{{\left| {{10}^{\frac{-{{L}_{q,m}^{t}}}{10}}} \right|}^{2}}}},
\end{align}
where ${{\beta }_{q,m,r}}=1$ represents that the channel $r$ is assigned to DT $m$. Otherwise, ${{\beta }_{q,m,r}}=0$. Moreover, ${{P}_{q,m,r}}$ is the transmission power. Hence, the processed number of tasks in the computation offloading mode is denoted as
\begin{align}
R_{n,m,r}^{t}={{B}_{n,m,r}}\log \left( 1+\frac{{{a}_{n,m,o}}{{P}_{n,m,r}}|{{10}^{\frac{-L_{n,m}^{t}}{10}}}|^{2}}{{{\sigma }^{2}}+{{I}_{n,m,r}}} \right),
\end{align}
where ${{B}_{n,m,r}}$ is the allocated channel bandwidth for DT $m$. ${{a}_{n,m,o}}=1$ indicates that DT $m$ offloads the task to the LEO $o$ and ${{\sigma }^{2}}$ is channel noise power. Consequently, the processed number of bits is represented as
\begin{align}
D_{n,m}^{t2}=R_{n,m,r}^{t}T.
\end{align}
\subsection{Blockchain Verification Model}
\ \ Since multiple DTs cannot share task data with each other, we utilize the blockchain technology to strengthen data privacy and prevent data tampering. Hence, each block records related model parameters and these information is verified via corresponding DTs. Subsequently, the privacy protection overhead is divided into parameters aggregation, transmission, and verification parts. Specifically, the parameters aggregation overhead is denoted as
\begin{align}
C1=\frac{{\left|{{W}_{m}}\right|}}{{{f}_{MBS}}}, \end{align}
where $\left| {{W}_{m}} \right|$ and ${{f}_{MBS}}$ are corresponding model training parameters and CPU cycle frequency from the MBS. Next, the parameters transmission overhead is denoted as
\begin{align}
C2=\delta{{\log }_{2}}N\frac{{|{{W}_{m}}|}}{{{r}_{up}}},
\end{align}
where $\delta$ is the model transmission factor and ${{r}_{up}}$ is uplink transmission rate from the MBS to the LEO. Finally, the parameters verification overhead is represented as
\begin{align}
C3=\delta {{\log }_{2}}MN\frac{{{S}_{B}}}{{{r}_{down}}}+\underset{\{m\}}{\mathop{\max }}\,\left\{ \frac{{{S}_{B}}}{f_{n,m}^{t}} \right\},
\end{align}
where ${{S}_{B}}$ is the size of blockchain and ${{r}_{up}}$ is data downloading rate from the LEO to the MBS. Finally, the total blockchain privacy protection overhead is calculated as
\begin{align}
{{C}_{SBC}}=C1+C2+C3.
\end{align}
\subsection{Problem Formulation}
\ \ In this section, we formulate the two-stage Stackelberg game model between the cloud servers and the terrestrial users. As terrestrial users can be regarded as DTs projected to the MBS, the cloud providers need to set the price to earn more profits after collecting the CPU cycle frequency $f_{n,m}^{t}$. Hence, the corresponding cloud servers profits are represented as
\begin{align}
& P1: \ \underset{{{\lambda }_{n,m}}}{\mathop{\max }}\,\sum\limits_{n=1}^{N}{\sum\limits_{m=1}^{M}{{{\lambda }_{n,m}}f_{n,m}^{t}-cf_{n,m}^{t}}} \\
& s.t. \ \ \ \ \ {{\lambda }_{n,m}}\ge 0,
\end{align}
where ${\lambda }_{n,m}$ indicates the cloud servers' price and $c$ is one unit electronic consumption. Next, for massive DTs $m$, the network throughput is calculated after knowing the pricing policy from cloud servers, which is represented as
\begin{align}
F1=\underset{T\to \infty }{\mathop{\lim }}\,\frac{1}{T}\sum\limits_{t=1}^{T}{\alpha _{n,m}^{t}D_{n,m}^{t1}+(1-\alpha _{n,m}^{t})D_{n,m}^{t2}}.
\end{align}
Next, the privacy overhead and price loss are denoted as
\begin{align}
F2={{C}_{SBC}}-{{\lambda }_{n,m}}f_{n,m}^{t}.
\end{align}
Furthermore, the total service demands for users are represented as
\begin{footnotesize}
\begin{align}
& P2: \underset{\{f_{n,m}^{t},\alpha _{n,m}^{t},{{a}_{n,m,o}},{{S}_{B}}\}}{\mathop{\max }}\,\{F1-F2\}\\
& \ \ \ \ \ s.t.\ \ \ \ \ \ \ \ \ \ \ \ \ \ \ \ \ \frac{f_{n,m}^{t}}{w}\le Q_{n,m}^{t}, \\
& \ \ \ \ \ \ \ \ \ \ \ \ \ \ \ \ \ \ \ \ \ \ \ \ \ \ \ {{I}_{n,m,r}}\le {{I}_{\max }}, \\
& \ \ \ \ \ \ \ \ \ \ \ \ \ \ \ \ \ \ \ \ \ \ \ \ \ \ \ \alpha _{n,m}^{t}\in \{0,1\}, \\
& \ \ \ \ \ \ \ \ \ \ \ \ \ \ \ \ \ \ \ \ \ \ \ \ \ \ \ \underset{T\to \propto }{\mathop{\lim }}\,\frac{1}{T}\sum\limits_{t=1}^{T}{E[Q_{n,m}^{t}]}<\propto , \\
& \ \ \ \ \ \ \ \ \ \ \ \ \ \ \ \ \ \ \ \ \ \ \ \ \ \ \ {{S}_{\min }}\le {{S}_{B}}\le {{S}_{\max }},
\end{align}
\end{footnotesize}where (16) represents that the processed number of tasks cannot exceed the task queue $Q_{\text{n},m}^{t}$, (17) means that the channel interference should be less than maximum ${{I}_{\max }}$, $\alpha _{n,\text{m}}^{t}$ is corresponding offloading decision, and (19) indicates that the long-term task queue for each DT is limited. Finally, (20) denotes the block size.

\section{Algorithm Design in SAG-DT integrated Blockchain Network}

\subsection{Lyapunov-based Problem Transformation}
\begin{figure*}[!t]
\centering{}
\setlength{\abovecaptionskip}{0.5cm}
\includegraphics[width=16cm, height=8cm]{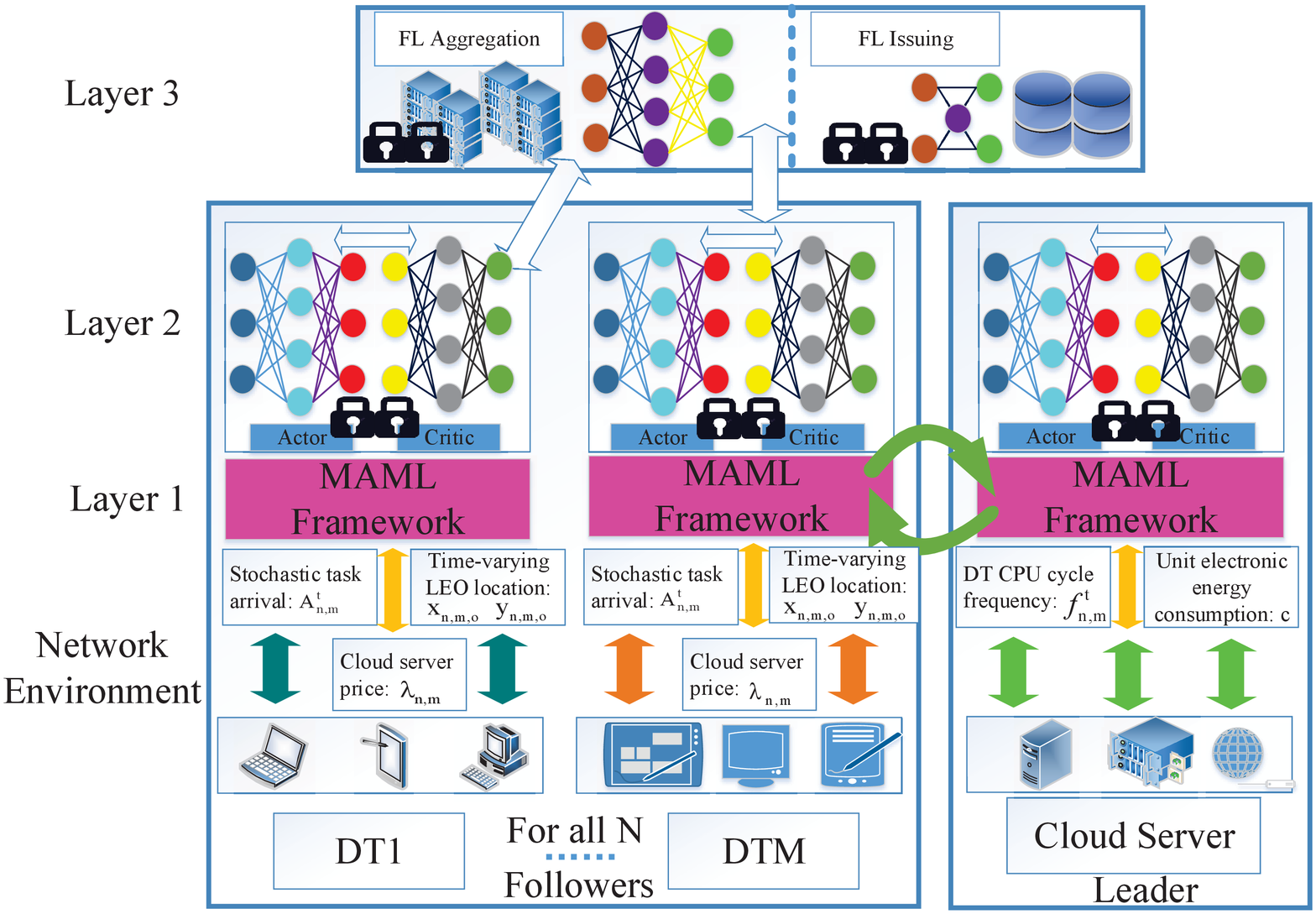}
\caption{The proposed three-layer MAML-MADFRL framework.}
\setlength{\belowdisplayskip}{-0.1cm}
\label{The proposed three-layer MAML-MADFRL algorithm framework}
\end{figure*}
\ \ In terms of P1 and P2, when the network provider sets the service price $\lambda_{n,m}$, multiple DTs the make optimal task decision-makings in terms of cloud server pricing, which can be regarded as a two-stage Stackelberg game process. Hence, we need to explore the optimal DTs service demands according to servers pricing. However, P2 indicates the variables coupling between long-term task queue and short-term computation offloading as well as privacy protection, which make it hard to decouple corresponding optimization variables. Subsequently, we introduce Lyapunov stability theory to further transform multiple time slots problem into a single time slot subproblem. Meanwhile, we denote the virtual task queue $\{\vec{Q}(t)\}_{n,m}^{t}$ to fulfill the long-term task queue constraint. The corresponding Lyapunov function and drift function are denoted as
\begin{align}
L(\vec{Q}(t))=0.5{{{(Q_{n,m}^{t})}^{2}}},
\end{align}
\begin{align}
\Delta L(\vec{Q}(t))=\mathbb{E}\left\{ L(\vec{Q}(t+1))-L(\vec{Q}(t))|\vec{Q}(t) \right\}.
\end{align}
In terms of (21) and (22), we can decompose the long-term constraint (19) and further derive the minimum values of drift-plus-penalty algorithm. Moreover, we denote the virtual queue backup $Q_{n,m}^{t}$ as
\begin{align}
Q_{n,m}^{t+1}=\max \left\{ Q_{n,m}^{t}-D_{n,m}^{t1/t2}+A_{n,m}^{t},0 \right\},
\end{align}
where $D_{n,m}^{t1/t2}=\alpha_{n,m}^{t}D_{n,m}^{t1}+(1-\alpha _{n,m}^{t})D_{n,\text{m}}^{\text{t2}}$.

We further simplify (23) as
\begin{align}
& {{(Q_{n,m}^{t+1})}^{2}}={{(Q_{n,m}^{t})}^{2}}+2Q_{n,m}^{t}(A_{n,m}^{t}-D_{n,m}^{t1/t2})+{{(A_{n,m}^{t}-D_{n,m}^{t1/t2})}^{2}}.
\end{align}
Moreover, we execute summation for (24)
\begin{align}
& \ \ \ \ \ \ \ \ \ 0.5{{{\left( Q_{n,m}^{t+1} \right)}^{2}}-}0.5{{{\left( Q_{n,m}^{t} \right)}^{2}}}={Q_{n,m}^{t}(A_{n,m}^{t}-D_{n,m}^{t1/t2})+0.5{{{(A_{n,m}^{t}-D_{n,m}^{t1/t2})}^{2}}}}.
\end{align}
Next, we derive the Lyapunov drift as
\begin{align}
& \vartriangle L(\vec{Q}(t))=\mathbb{E}\left\{ L(\vec{Q}(t+1))-L(\vec{Q}(t)) \right\} =0.5{{{\left( A_{n,m}^{t}-D_{n,m}^{t1/t2} \right)}^{2}}+{Q_{n,m}^{t}\left( A_{n,m}^{t}-D_{n,m}^{t1/t2} \right)}} \nonumber\\
& \le X+{Q_{n,m}^{t}\left( A_{n,m}^{t}-D_{n,m}^{t1/t2} \right)},
\end{align}
where
\begin{align}
& 0.5{{{\left( A_{n,m}^{t}-D_{n,m}^{t1/t2} \right)}^{2}}\le 0.5{{{\left( A_{n,m}^{t} \right)}^{2}}+{{\left( D_{n,m}^{t1/t2} \right)}^{2}}}}\le 0.5{{{\eta }_{n,m}}}+{{\left( D_{n,m}^{t1/t2/\max} \right)}^{2}}=X,
\end{align}
where $D_{n,m}^{t1/t2/\max }$ is the maximum of $D_{n,m}^{t1/t2}$. Moreover, it denotes the maximum local DT execution and offloading execution.

Furthermore, the drift-plus-penalty equation is calculated as
\begin{align}
\Theta \left( \vec{Q}(t) \right)=\Delta L\left( \vec{Q}(t) \right)-V\mathbb{E}\{F\},
\end{align}
where $V$ is the corresponding Lyapunov control parameter and
\begin{align}
& F=\alpha _{n,m}^{t}D_{n,m}^{t1}+(1-\alpha _{n,m}^{t})D_{n,m}^{t2}-{{C}_{SBC}}-{{\lambda }_{n,m}}f_{n,m}^{t}.
\end{align}
Hence, the original optimization P2 for all DTs is transformed into
\begin{align}
& P{{2}^{'}}:\ \ \ \ \underset{\{\forall m\}}{\mathop{\max }}\ \ \,{Q_{n,m}^{t}D_{n,m}^{t1/t2}+VF} \\
& s.t. \ \ \ \ \ \ (16),(17),(18),(20).\nonumber
\end{align}
Specifically, we transform the long-term multi-slot optimization problem into a single-slot optimization problem, which utilizes the proposed Lyapunov drift-plus-penalty algorithm to maximize the number of processed bits in terms of minimum privacy overhead and pricing strategy from cloud servers. Subsequently, the specific Lyapunov-based problem transformation is shown in Algorithm 1.

\textbf{Remark 1}: For the Lyapunov-based problem transformation, we further decompose the long-term multi-slot problem into a single-slot optimization problem for all followers. After the Lyapunov-based problem transformation, it is not necessary to know future system state information and probability distribution of random events, which is regarded as a model-free learning. Moreover, the proposed Lyapunov-based transformation mechanism and MAML-MADFRL algorithm can satisfy the constraint in (19), which are verified in Section V.
\begin{algorithm}
\setstretch{1}
  \caption{Lyapunov drift-plus-penalty algorithm}
  \label{alg1}
  \begin{algorithmic}[1]
   \REQUIRE~~\\
   The original multi-slot optimization problem P2 for all followers.
   \ENSURE ~~\\
   The single-slot optimization problem $P{{2}^{'}}$.
   \STATE Define the virtual task queue $Q_{n,m}^{t}$;\\
   \STATE Compute the Lyapunov function $L\left( \vec{Q}(t) \right)$ and Lyapunov drift function $\Delta L\left( \vec{Q}(t)\right)$.\\
   \STATE Derive the Lyapunov drift from (23) to (27);\\
   \STATE Obtain the drift-plus-penalty equation in (28);\\
   \STATE Output the single-slot optimization problem $P{{2}^{'}}$;\\
  \end{algorithmic}
\end{algorithm}
\subsection{MAML-MADFRL Algorithm Design}
\ \ As shown in Fig. \ref{The proposed three-layer MAML-MADFRL algorithm framework}, we envision a three-layer MAML-MADFRL framework, which consists of a MAML fast adaptation module and a MADFRL optimization module. Moreover, the framework includes multiple followers (i.e., DTs) and one cloud server (leader). Specifically, multiple DTs can adapt to stochastic task arrival ($A_{n,m}^{t}$), time-varying LEO location ($x_{n,m,o}$, $y_{n,m,o}$), and cloud server price (${{\lambda }_{n,m}}$), which can further obtain the optimal CPU cycle frequency ${{f}_{n,m,t}}$, task offloading decision-making ${{\alpha }_{n,m,t}}$, channel selection ${{a }_{n,m,o}}$, and block size ${{S}_{B}}$. Meanwhile, the leader can set the cloud server price (${{\lambda }_{n,m}}$) in terms of ${{f}_{n,m,t}}$ and unit electronic energy consumption $c$, which in turn helps each DT make corresponding execution actions.

Subsequently, we utilize the proposed FL aggregation and FL issuing policies to further accelerate model convergent rate and protect user privacy. More importantly, the FL aggregation mechanism adjusts corresponding DT actor network parameters in terms of task weight and time-varying LEO locations. After finishing FL aggregation, the corresponding model parameters are issued to each DT, which can further achieve privacy protection and security verification. Next, we introduce related MAML fast adaptation algorithms and MADFRL optimization mechanisms.

\subsubsection{MAML Fast Adaptation Algorithm}
The goal of MAML is to rapidly learn new tasks from small batches of samples, which are more efficient than learning from scratch. Moreover, MAML can collect historical experience from past tasks to rapidly adapt to new tasks. Assuming that these old tasks for meta-training and new tasks for meta-testing are subject to the same basic distribution $p(\Gamma)$, there are some common characteristics among different tasks. For conventional DRL scenarios, the aim is to minimize the loss function ${{L}_{\Gamma}}$ for specific tasks $\Gamma$. However, MAML can learn corresponding network weight parameters ${{w}^{'}}={{u}_{\varphi }}\left( D_{\Gamma }^{tr},w \right)$, which can utilize small batch of samples to rapidly adapt to new tasks $\Gamma$. Hence, MAML-RL problem is represented as
\begin{align}
& \underset{w}{\mathop{\min }}\,\ \ \ \ \ {{\mathbb{E}}_{\Gamma \tilde{\ }p(\Gamma)}}\left[ L\left( D_{\Gamma }^{test},{{w}^{'}} \right) \right] \\
& s.t.\ \ \ \ \ {{w}^{'}}={{\mu }_{\varphi }}\left( D_{\Gamma }^{tr},w \right),
\end{align}
where $D_{\Gamma}^{tr}$ and $D_{\Gamma}^{test}$ represent training task samples and testing task samples from $p\left(\Gamma\right)$ and $L\left( D_{\Gamma }^{test},{{w}^{'}} \right)$ means the testing loss function from new network weight parameters ${{w}^{'}}$ in terms of testing task samples. Next, MAML process is divided into two parts, i.e., the inner loop and outer loop. For the inner loop, we sample new training tasks to update network weight parameters ${{w}^{'}}$, which are utilized to test model in the outer loop. Subsequently, we regard testing task samples error as loss function to retrain network model in the outer loop. Noting that the model is an initial model for the inner loop and it only generates testing error for training tasks in the inner loop. Finally, MAML is responsible for updating initial model. The detailed MAML algorithm is shown in Algorithm 2.

\textbf{Remark 2}: MAML is based on gradient descent, and ${{w}^{'}}={{u}_{\varphi }}\left( D_{\Gamma }^{tr},w \right)$ is updated via several gradient descent steps to obtain better network performance gains for new testing tasks. The goal of MAML is to train and obtain model initial parameters, which maximizes the new tasks performance via several gradient update steps from small batches of samples. Although these update principles are fixed, a set of well-optimized neural network parameters followed with several gradient update steps are utilized to generalize new testing tasks.
\begin{algorithm}
\setstretch{1}
  \caption{MAML for MADFRL}
  \label{alg1}
  \begin{algorithmic}[1]
   \REQUIRE~~\\
   Followers: stochastic tasks arrivals $p\left(\Gamma\right)$ for followers, time-varying LEO location ${{x}_{n,m,o}}$ and ${{y}_{n,m,o}}$, cloud server price ${{\lambda }_{n,m}}$;\\
   Leader: DT CPU cycle frequency $f_{n,m}^{t}$ and unit electronic energy consumption $c$;\\
   Learning rate: $\alpha $ and $\beta$;\\
   \ENSURE ~~\\
   Neural network parameters ${{w}^{'}}$;\\
   \STATE Randomly initialize neural network parameters $w$;\\
   \WHILE {Not finished}
   \STATE Extract small batches of states $\Gamma$ from all followers and the leader;\\
   \FOR {all $\Gamma$}
   \STATE Extract $K$ MDP trajectories $\Omega =\left\{ S_{n,m}^{'}|{{S}_{n,m}},{{A}_{n,m}},{{R}_{n,m}} \right\}$ utilizing $w$ in $\Gamma$;\\
   \STATE Compute ${{\nabla }_{w}}{{L}_{\Gamma }}(w)$ utilizing $\Omega$ and ${{L}_{\Gamma }}(w)$;\\
   \STATE Update neural network parameters via gradient descent: ${{w}^{'}}=w-\alpha {{\nabla }_{w}}{{L}_{\Gamma }}\left( w \right)$;\\
   \STATE Continue to extract MDP trajectories $\Omega^{'} =\left\{ S_{n,m}^{'}|{{S}_{n,m}},{{A}_{n,m}},{{R}_{n,m}} \right\}$ using ${{w}^{'}}$;\\
   \ENDFOR
   \STATE Update $w=w-\beta {{\nabla }_{w}}\sum\limits_{\Gamma }^{{}}{{{L}_{\Gamma }}\left( {{w}^{'}} \right)}$ utilizing $\Omega^{'}$;\\
   \ENDWHILE
  \end{algorithmic}
\end{algorithm}

\subsubsection{MADFRL Optimization Mechanism}
For the MAML algorithm, it helps network model obtain the optimal policy of new tasks from small samples via several gradient descent steps. Subsequently, we explore the MADFRL algorithm principle in Layer 2 and Layer 3. In Layer 2, we employ actor-critic network model to execute task offloading as well as resource allocation and adapt to dynamic network environment. Specifically, for multiple followers from $1$ to $M$, we utilize multi-agent actor-critic network structure to process corresponding stochastic task arrival $A_{n,m}^{t}$, time-varying LEO location ${{x}_{n,m,o}}$ as well as ${{y}_{n,m,o}}$, and cloud server price ${{\lambda }_{n,m}}$, which cannot cause any information exchange among multiple followers to protect followers' privacy. Moreover, the proposed multi-agent actor-critic framework consists of a actor neural network and a critic neural network. The specific state, action and reward function for followers are defined as follows.

{\em State Space}: At the beginning of time slot $t$, each follower $m$ in the ${{n}_{th}}$ MBS receives the local state ${{S}_{n,m}}=\left\{ A_{n,m}^{t},{{x}_{n,m,o}},{{y}_{n,m,o}},{{\lambda }_{n,m}}\right\}$, and ${{S}_{n,m}}$ is not allowed to interact among multiple followers because of privacy protection. Meanwhile, the network state is constant in a single time slot $t$ but varies across different time slots.

{\em Action Space}: For the multi-agent actor-critic framework, each follower makes following four actions via actor neural network, i.e., CPU cycle frequency $f_{n,m}^{t}$, channel selection $\alpha _{n,m,t}^{{}}$, task offloading action ${{a}_{n,m,o}}$, and block size ${{S}_{B}}$. Hence, the action space for each follower is denoted as
${{A}_{n,m}}=\left( f_{n,m}^{t},\alpha _{n,m,t}^{{}},{{a}_{n,m,o}},{{S}_{B}} \right)$, which is executed via actor neural network.

{\em Reward Function}: As we intend to maximize the processed task numbers while minimizing the privacy protection overhead and frequency consumption, each follower needs to maximize own reward function. Subsequently, the instant reward is represented as
\begin{align}
R_{n,m}=\sum\limits_{n=1}^{N}{\sum\limits_{m=1}^{M}{Q_{n,m}^{t}D_{n,m}^{t1/t2}+VF}}.
\end{align}
Hence, in the training and testing phase, each follower can maximize the reward function from small batches of samples.

{\em MDP Transition Process}: For the blockchain-aided SGIDTN model, it is hard to find a fixed transition policy to cover network states. Consequently, we can use $\Omega =\left\{ S_{n,m}^{'}|{{S}_{n,m}},{{A}_{n,m}},{{R}_{n,m}} \right\}$ to further represent the state transition between followers and network environment.

Similar to followers, the state space, action space and reward function for the leader are  be represented as follows.

The leader needs to adjust own price policy ${{\lambda }_{n,m}}$ according to DT CPU cycle frequency $f_{n,m}^{t}$ and unit electronic energy consumption $c$. Hence, the corresponding state space for the leader is ${{S}_{L}}=\left\{ f_{n,m}^{t},c \right\}$. Next, after obtaining the state space, each actor neural network generates network serving price ${{A}_{L}}=\left\{{{\lambda }_{n,m}}\right\}$. Moreover, the related reward function is denoted as
${{R}_{L}}=\sum\limits_{n=1}^{N}{\sum\limits_{m=1}^{M}{{{\lambda }_{n,m}}f_{n,m}^{t}-cf_{n,m}^{t}}}$. After multiple iterations and training, the leader will obtain the maximum profits. Subsequently, the MDP transition policy for the leader is denoted as $\Psi =\{S_{L}^{'}|{{S}_{L}},{{A}_{L}},{{R}_{L}}\}$. The specific MADFRL algorithm for multiple followers is illustrated in Algorithm 3.

\begin{algorithm}
\setstretch{1}
  \caption{The specific MADFRL algorithm}
  \label{alg1}
  \begin{algorithmic}[1]
  \REQUIRE~~\\
   Followers: $A_{n,m}^{t}$; ${{x}_{n,m,o}}$; ${{y}_{n,m,o}}$ and ${{\lambda }_{n,\text{m}}}$.
   Leader: $f_{n,m}^{t}$ and $c$.
   \ENSURE ~~\\
   Actor network weight parameters.
  \FOR {each MBS $n \in \{1,2,..,N\}$}
  \FOR {each follower $m \in \{1,2,..,M\}$}
  \STATE Initialize actor network parameters $w$; critic network parameters $Q({{S}_{n,m}},{{A}_{n,m}})$;
   \ENDFOR
   \ENDFOR
    \FOR {each $t \in \{1,2,..,T\}$}
    \FOR {each $n \in \{1,2,..,N\}$}
   \FOR {each $m \in \{1,2,..,M\}$}
   \STATE Execute the current task decision-making ${{A}_{n,m}}$;
   \STATE Obtain the reward function ${{R}_{n,m}}$;
   \STATE Transfer to next states $S_{n,m}^{'}$;
   \STATE Input ${{A}_{n,m}}$ and ${{S}_{n,m}}$ to critic networks and obtain $Q\left( {{S}_{n,\text{m}}},{{A}_{n,m}} \right)$;
   \STATE Update critic network parameters via ${{\sum\limits_{{}}^{{}}{\left( {{R}_{n,m}}+\gamma Q\left( {{S}_{n,m}},{{A}_{n,m}} \right) \right)}}^{2}}$;
   \STATE Update actor network via $-Q\left( {{S}_{n,m}},{{A}_{n,m}} \right)$
   \STATE Record the state transitions $\Omega =\left\{ S_{n,m}^{'}|{{S}_{n,m}},{{A}_{n,m}},{{R}_{n,m}} \right\}$;
   \ENDFOR
   \STATE Compute the corresponding reward function ${{R}_{n,m}}$;
   \STATE Output actor network parameters ${{Z}_{n,m}}$ and transmit them to Layer 3 in order to execute FL aggregation and issuing;
   \ENDFOR
   \ENDFOR
  \end{algorithmic}
\end{algorithm}
\subsection{Federated Aggregation and Parameters Issuing Mechanism}

\ \ After each actor network outputs the offloading policy, we need to appraise these actor network actions to further adjust neural network parameters. However, the information exchange among multiple followers to prevent user privacy leakage. Hence, we propose the FL aggregation and issuing mechanism to centrally update actor network parameters and distributively issue them to each follower. Specifically, the FL aggregation and issuing module belong to Layer $3$ for the proposed MAML-MADFRL framework and Layer 3 can collect all actor network parameters from followers in order to average followers' network parameters via different tasks size and time-varying location weights, which can then issue these network weight parameters to each follower. As actor network has lightweight network parameters \cite{zhu2021federated}, it is beneficial to decrease parameters transmission overhead and improve communication efficacy in terms of large data volume. Furthermore, all followers receive the global update model $Z(t)$ from the leader according to computation offloading policy. Meanwhile, each follower can obtain actor neural network model ${{Z}_{n,m}}(t)$ in terms of corresponding state $S{}_{n,m}$. Hence, each follower uploads the new model to the leader, which is represented as
\begin{align}
{{I}_{n,m}}(t)=Z(t)-{{Z}_{n,m}}(t).
\end{align}
\ \ Next, the leader receives the uploaded model and further optimizes them via the proposed federated aggregation policy, which are generally defined as
\begin{align}
Z(t+1)=Z(t)+uI(t),
\end{align}
where $u$ is the FL aggregation learning rate and $I(t)$ is represented as
\begin{align}
I(t)={\frac{\left| {{D}_{n,\text{m}}} \right|+\left| {{L}_{\text{n},m}} \right|}{\left| {{D}_{total}} \right|+\left| {{L}_{total}} \right|}}{{I}_{n,m}}(t),
\end{align}
where ${{D}_{n,m}}$ and ${{D}_{total}}$ are task size of each follower and all task sets from followers, respectively. Meanwhile, ${{L}_{n,m}}$ is the relative position distance from LEO to ground follower and ${{L}_{total}}$ is the sum of total distance for all followers. As illustrated above, larger task bits and relative location distance mean more model update ratio. The specific federated aggregation and issuing mechanism are shown in Algorithm 4.
\begin{algorithm}
\setstretch{1.15}
  \caption{The FL aggregation mechanism}
  \label{alg1}
  \begin{algorithmic}[1]
  \REQUIRE~~\\
   Actor network parameters ${{Z}_{n,m}}(t)$;
   \ENSURE ~~\\
    Issue actor network weight for each follower;
   \FOR {each $n \in \{1,2,..,N\}$}
   \FOR {each $m \in \{1,2,..,M\}$}
   \STATE Upload respective actor network parameters ${{Z}_{n,m}}(t)$;
   \STATE Execute the FL aggregation and issuing mechanism from (34) to (36);
   \ENDFOR
   \ENDFOR
   \STATE Obtain the optimal actor network parameters.
  \end{algorithmic}
\end{algorithm}

\textbf{Remark 3}: For multiple followers, we execute FL aggregation and issuing mechanism to weigh actor network parameters, which aims to protect data privacy. Meanwhile, each actor network parameter is issued to corresponding followers in terms of different tasks and location ratios. Hence, larger task and location ratio mean more parameters weight issuing for each follower, which help maximize the processed number of task bits in terms of minimizing privacy overhead and cloud server cost.

\section{Performance Analysis}
\ \ For this section, we explore the corresponding privacy protection-based blockchain verification mechanism for followers and leader, algorithm complexity analysis, long-term task queue constraints and MAML-MADFRL algorithm convergent rate.

\subsection{Transaction Verification}
\begin{figure}[!h]
\centering{}
\setlength{\abovecaptionskip}{0.5cm}
\includegraphics[width=12cm, height=8cm]{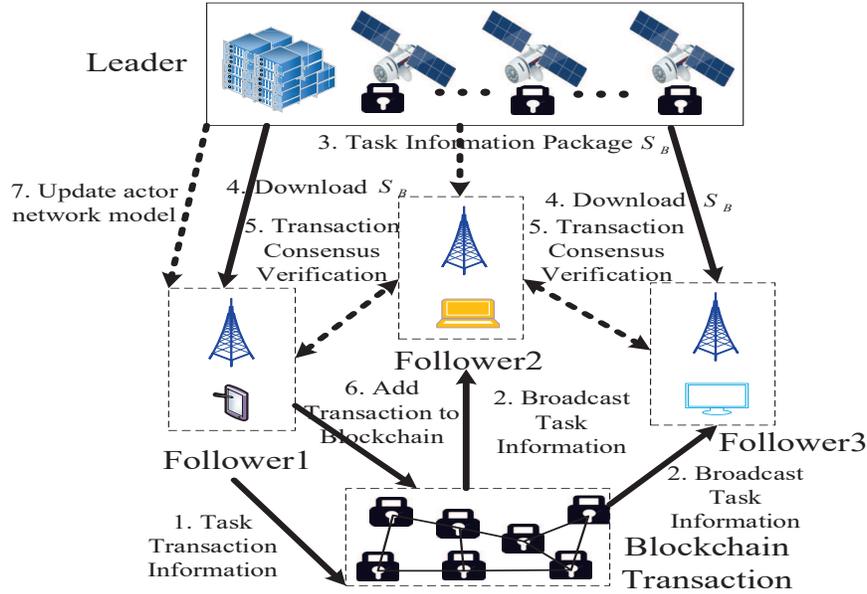}
\caption{Privacy protection-based blockchain mechanism for followers and the leader.}
\setlength{\belowdisplayskip}{0.1cm}
\label{The blockchain model}
\end{figure}
\ \ The complete privacy protection based-blockchain verification mechanism is shown in Fig. \ref{The blockchain model}, which can be divided into several procedures. First, follower $1$ generates the task transaction information stored in the blockchain. Next, these task information is broadcasted to other followers and transmitted to the leader. Subsequently, the leader collects these task information and packages them into ${{S}_{B}}$. Moreover, each follower in the MBS starts to download block ${{S}_{B}}$ and the transaction consensus verification is achieved via delegated stakes proof protocols \cite{li2022efficient}. After the task transaction process is passed via other followers, the coin rewards are returned back to initiators. It it suffers from fraud or tampering, the task transaction is closed. Finally, the verified transaction is added to blockchain and the follower updates actor network parameters.
\subsection{Algorithm Complexity Analysis}
\ \ In this subsection, we derive computational complexity of the proposed four algorithms, i.e., the Lyapunov drift-plus-penalty algorithm, MAML for MADFRL algorithm, MADFRL algorithm, and FL aggregation mechanism. Specifically, as the Lyapunov drift-plus-penalty algorithm transforms the original problem P2 into $P{{2}^{'}}$, the computational complexity of the Lyapunov drift-plus-penalty algorithm is $O(1)$. Next, the computation complexity of MAML for MADFRL can be calculated as $O\left( G+G\Gamma K \right)$, where $G$ is the number of task execution. Subsequently, while executing the task offloading and resource allocation, the MADFRL algorithm complexity is represented as $O\left[ MN\left( 1+T\Gamma K \right) \right]$. Finally, after each follower transfers these network parameters to the leader, the algorithm complexity for FL aggregation mechanism is represented as $O\left( MN \right)$.
\subsection{Convergence and Constraints Demonstration}
\ \ In this subsection, we discuss the asymptotic convergent performance of the proposed MAML-MADFRL algorithm and validate the long-term task queue constraints. Subsequently, we introduce related preliminary knowledge about Lyapunov optimization for verifying the convergent performance and task queue constraints, which is regarded as \textbf{W-only} policy independent on task queue backlogs.

\textbf{W-only policy}: The policy only relies on observed network states to choose optimal control actions, i.e., for any observed network states ${{S}_{n,m}}$, W-only policy can choose actions according to certain probability distribution, which is represented as
\begin{align}
{{\alpha }^{*}}(t)=\arg \max \Pr \left( \alpha (t)|{{S}_{n,m}}(t) \right).
\end{align}
Subsequently, the W-only policy does not rely on any queue states. In terms of Lyapunov stability theory \cite{neely2010stochastic}, when the original problem is feasible, there must exist the optimal \textbf{W-only policy}. Hence, the \textbf{W-only policy} provides a method to form the optimal solution.

\subsubsection{Time average penalty analysis} We further derive the asymptotic optimal performance of the proposed MAML-MADFRL algorithm via introducing the \textbf{W-only policy}, which is represented as
\begin{align}
& \Theta \left( \vec{Q}\left( t \right) \right)=\Delta L\left( \vec{Q}(t) \right)-V\mathbb{E}\{F\}\le X+Q_{n,m}^{t}\left( A_{n,m}^{t}-D_{n,m}^{t1/t2}\left( \alpha  \right) \right)-V\mathbb{E}\{F\left( \alpha  \right)\}\\
&=X+Q_{n,m}^{t}A_{n,m}^{t}-Q_{n,m}^{t}D_{n,m}^{t1/t2}\left( \alpha  \right)-V\mathbb{E}\{F\left( \alpha  \right)\} \nonumber\le X+Q_{n,m}^{t}A_{n,m}^{t}-Q_{n,m}^{t}D_{n,m}^{t1/t2}\left( \beta  \right)-V{{S}^{*}},\nonumber
\end{align}
where $\alpha$ is the original action decisions via the MAML-MADFRL algorithm and drift-plus-penalty algorithm. Moreover, $\beta$ is the optimal \textbf{W-only policy} and ${{S}^{*}}$ is the optimal solution for P2. Meanwhile, as the optimal \textbf{W-only policy} $\beta$ can enable the maximum number of task processed, $D_{n,m}^{t1/t2}\left( \beta  \right)$ is equal to $A_{n,m}^{t}$. Hence, (38) can be further simplified as
\begin{align}
\Delta L\left( \vec{Q}(t) \right)-V\mathbb{E}\{F\}\le \hat{X}-Q_{n,m}^{t}D_{n,m}^{t1/t2}(\beta )-V{{S}^{*}},
\end{align}
where $\hat{X}=X+Q_{n,m}^{t}A_{n,m}^{t}$. Next, we take the summation for both sides of (39) from $1$ to $T$, which is derived as
\begin{align}
& \left( X-V{{S}^{*}} \right)T\ge \sum\limits_{t=1}^{T}{\mathbb{E}\left[ \Delta L\left( \vec{Q}(t)-V\mathbb{E}\{F\} \right)|\vec{Q}(t) \right]}\ge \mathbb{E}\left[ L\left( \vec{Q}(T) \right) \right]-V\sum\limits_{t=1}^{T}{\mathbb{E}\left\{ F|\vec{Q}(t) \right\}} \nonumber \\
& \ge -V\sum\limits_{t=1}^{T}{\mathbb{E}\{F|\vec{Q}(t)\}}.
\end{align}
Finally, the time average penalty is represented as
\begin{align}
\underset{T\to \infty }{\mathop{\lim }}\,\sum\limits_{t=1}^{T}{\frac{1}{T}}\mathbb{E}\left\{ F|\vec{Q}(t) \right\}\le {{S}^{*}}-\frac{{\hat{X}}}{V}.
\end{align}
\subsubsection{Time average queue size} First, we assume the \textbf{W-only policy} $\eta $ (not the optimal policy), which is represented as
\begin{align}
D_{n,m}^{t1/t2}\left( \eta  \right)\ge \varsigma,
\end{align}
where $\varsigma>0$. Meanwhile, assuming that $F$ is a bounded function, which is represented as
\begin{align}
{{S}_{\min }}\le F\le {{S}_{\max }}.
\end{align}
Hence, the time average queue size is derived as
\begin{align}
& \Delta L\left( \vec{Q}(t) \right)-V\mathbb{E}\{F\}\le X+Q_{n,m}^{t}A_{n,m}^{t}-Q_{n,m}^{t}D_{n,m}^{t1/t2}\left( \eta  \right)-V\mathbb{E}\left\{F\right\} \nonumber\\
& \Delta L\left( \vec{Q}(t) \right)-V{{S}_{\max }}\le \hat{X}-V{{S}_{\min }}-Q_{n,m}^{t}\varsigma.
\end{align}
Furthermore, we take the summation and limit for both sides of inequality, which is represented as
\begin{align}
&\mathbb{E}\left[L\left( \vec{Q}(T\right) \right]\le T*\hat{X}+\left( {{S}_{\max }}-{{S}_{\min }}\right)VT-\sum\limits_{t=1}^{T}{Q_{n,m}^{t}\varsigma}\\
&\underset{T\to \infty }{\mathop{\lim }}\,\frac{\sum\limits_{t=1}^{T}{\mathbb{E}\left[ Q_{n,m}^{t} \right]}}{T}\le \frac{\hat{X}+\left( {{S}_{\max }}-{{S}_{\min}} \right)V}{\varsigma}.
\end{align}
Because of $\mathbb{E}\left[ L\left( \vec{Q}(t) \right) \right]\ge 0$, the time average queue size is proved.

\textbf{Remark 4}: It means that we can adjust the $V$ to asymptotically approach the maximum ${{S}^{*}}$ and satisfy the time average queue size $Q_{n,m}^{t}$. Specifically, (41) shows that when $V$ is larger, the time average penalty for the proposed MAML-MADFRL algorithm is closer to the optimal solution ${S}^{*}$. Moreover, the convergent rate for the time average penalty is $O\left( \frac{1}{V} \right)$. However, (47) indicates that if we only increase the Lyapunov control parameter $V$, it can cost longer to fulfill the long-term queue constraints, because the convergent rate for the average queue size is represented as $O\left( \frac{1}{V} \right)$. Finally, we show detailed performance results for the Lyapunov control parameter $V$ in the simulation results section.

\section{Simulation Results and Performance Analysis}
\ \ In this section, we have executed massive simulation experiments to validate the proposed algorithm performance. First, we demonstrate the impact of the Lyapunov control parameter $V$ on processed number of bits and time average queue. Next, we compare the proposed MAML-MADFRL algorithm with other three advanced schemes, which demonstrate that it has better performance gains in terms of pricing profits, network throughput, and reducing channel interference. Finally, the proposed blockchain-aided verification mechanism can not only protect users' privacy, but also reduce privacy verification overhead. The specific simulation details are shown next.
\subsection{Parameters Settings}
\ \ In the blockchain-aided SGIDTNs scenario, we explore the proposed MAML-MADFRL algorithm performance via massive simulation experiments. Specifically, we set that $N=4$ MBS, $M=12$ users and $O=4$ LEOs. Next, the stochastic task arrivals $A_{n,m}^{t}$ are uniformly distributed as (10, 30) MB and the required CPU number of cycles $w$ is uniformly distributed at (2,000, 4,000) cycle/bit. Moreover, the horizontal distance ${{x}_{n,m,o}}$ and vertical distance ${{y}_{n,m,o}}$ for LEOs are uniformly distributed at (1,000, 2,000) KM and (500, 2,000) KM. Furthermore, the carrier frequency ${{f}_{c}}$ and light speed $c$ are set as $0.1*{{10}^{9}}$ cycle/bit  HZ and $3*{{10}^{8}}$ m/s. Meanwhile, the path loss $\varepsilon _{n,m,o}^{LoS}$ and $\varepsilon _{n,m,o}^{NLoS}$ are uniformly distributed at (0,1) and (10, 30), respectively. The noise power $\sigma$ is ${{10}^{-13}}$ W. For MBS, the CPU cycle frequency ${{f}_{MBS}}$ is $6*{{10}^{9}}$ cycle/bit and the uplink rate as downlink rate is $0.5*{{10}^{10}}$ bit/s and $1*{{10}^{10}}$ bit/s, respectively.
\cite{qiu2020networking}
\subsection{Comparison Schemes}
\ \ We compare the proposed Lyapunov-aided MAML-MADFRL learning framework to other three baseline methods, such as multi-agent random task offloading (MARTO), multi-agent greedy channel selection (MAGCS), multi-agent mean CPU cycle (MAMCC). Moreover, we demonstrate the performance gains for MAML-MADFRL framework in terms of corresponding network parameters including task queue, Lyapunov control parameter, unit energy consumption, network throughput as well as channel interference, cloud server profits and privacy overhead.
\begin{itemize}
\item MARTO: For each stochastic task, the agent can randomly choose proper task processing units, such as local processing or LEO computation. However, it is hard to find the optimal task offloading decisions.
\item MAGCS: In order to reduce the channel interference from other areas, each agent randomly selects the proper channel according to path poss. That is, each agent tends to choose the channel with minimized path loss.
\item MAMCC: This method allocates the computational resources equally for each agent. While processing the stochastic tasks, each agent obtains the equal CPU cycle frequency to maximize the network throughput and cloud server profits.
\end{itemize}

\begin{figure*}[!t]
\subfigure[The average network throughput versus Lyapunov \protect\\ control parameter.]{
\begin{minipage}[t]{0.5\linewidth}
\centering{}
\includegraphics[width=7cm,height=4cm]{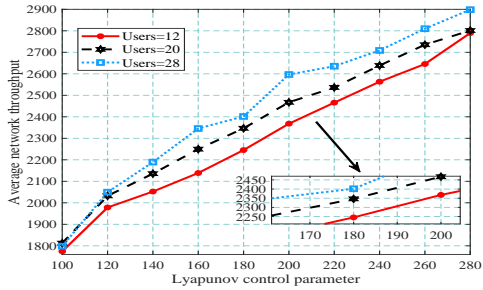}
\label{The average network throughput}
\end{minipage}%
}%
\subfigure[The time average task queue versus Lyapunov control \protect\\ parameter.]{
\begin{minipage}[t]{0.5\linewidth}
\centering{}
\includegraphics[width=7cm,height=4cm]{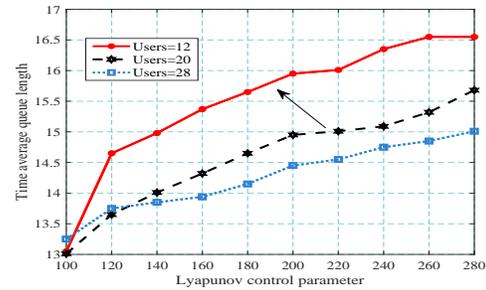}
\label{The time average queue length}
\end{minipage}%
}%
\centering{}
\caption{The impact of Lyapunov parameter on average network throughput and queue length.}
\end{figure*}

\subsection{Performance Evaluation}
\subsubsection{MAML-MADFRL Performance Gains}

Next, we show the performance gains of the proposed MAML-MADFRL algorithm in terms of different Lyapunov control parameters. As shown in Fig. \ref{The average network throughput}, we explore the impact of the Lyapunov control parameter on average network throughput according to different number of users. As the Lyapunov control parameter increases, it brings larger network throughput, which further demonstrates the time average penalty function. When there are more users, the proposed MAML-MADFRL method still stabilizes the network throughput because of meta learning and federated aggregation mechanism, which means that each user can process more task bits under complex network environment.

Next, as shown in Fig. \ref{The time average queue length}, we further explore the impact of Lyapunov control parameter on the time average queue length in terms of different number of ground users. For the sake of simplicity, we set the task size is 15 MB. Specifically, as the number of users increases, the time average queue length is smaller because the proposed MAML-MADFRL algorithm needs to adapt to more complex network environment and issues these parameters to more users. Moreover, when the Lyapunov control parameter increases, each user needs more time to converge to the optimal queue length, which further demonstrates the time average queue size.

\begin{figure*}[!t]
\subfigure[The cloud servers' profits versus unit energy consumption.]{
\begin{minipage}[t]{0.5\linewidth}
\centering{}
\includegraphics[width=7cm,height=4cm]{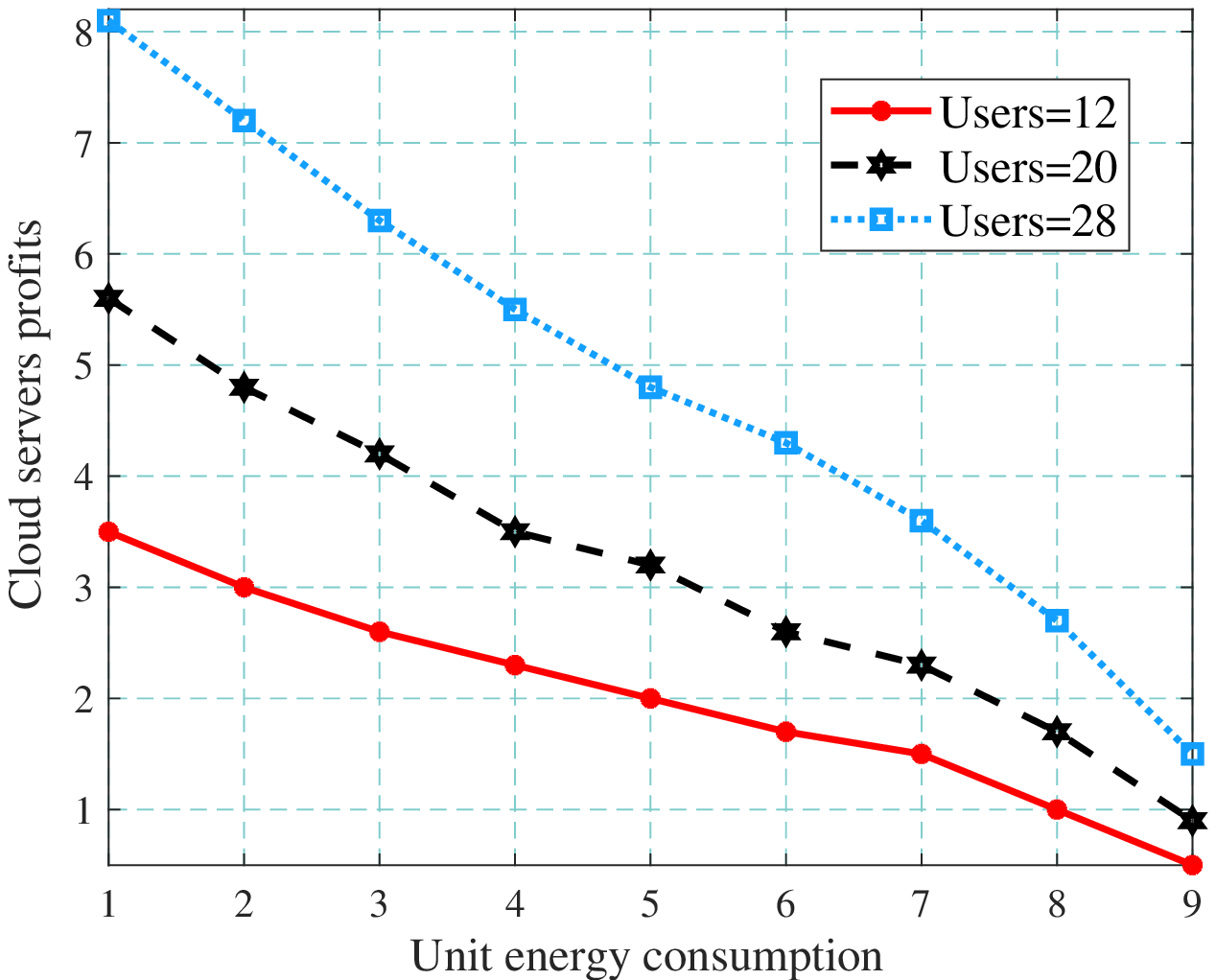}
\label{The cloud servers profits}
\end{minipage}%
}%
\subfigure[The mutual channel interference versus transmission power.]{
\begin{minipage}[t]{0.5\linewidth}
\centering{}
\includegraphics[width=7cm,height=4cm]{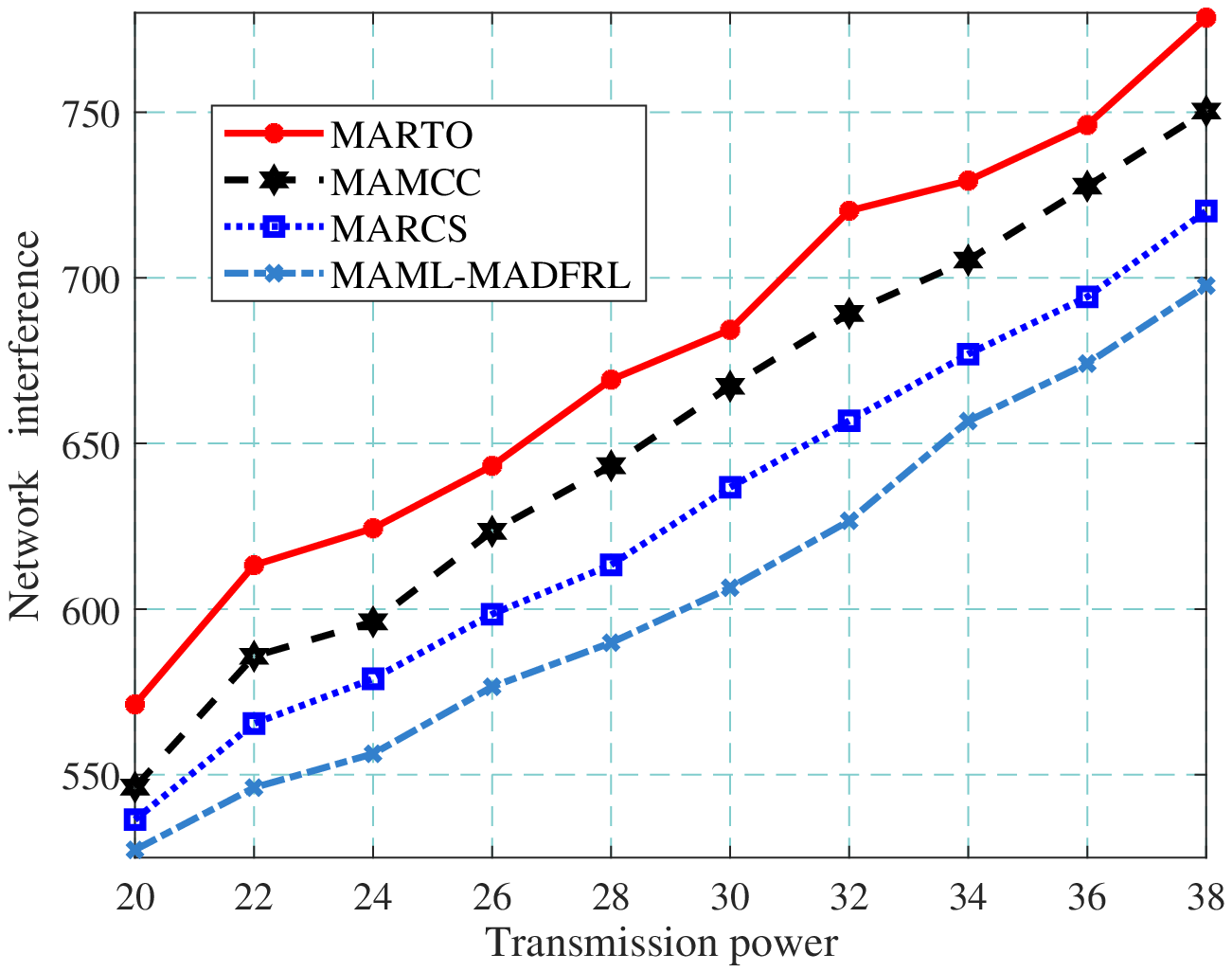}
\label{The mutual channel interference}
\end{minipage}%
}%
\centering{}
\caption{The cloud servers' profits and suffered network interference.}
\end{figure*}

Moreover, we explore the impact of unit energy consumption on cloud servers profits in terms of different number of users. As shown in Fig. \ref{The cloud servers profits}, when the unit energy consumption of cloud servers increases, it can bring lower servers profits because it serves these users relying on more energy consumption. Furthermore, as the number of users increases, the cloud server obtains higher profits. However, we need to find a tradeoff between the unit energy consumption and cloud server profits, which can reduce the energy consumption and improve corresponding profits.

\subsubsection{Algorithm Comparisons}
In this sequel, we compare our proposed MAML-MADFRL algorithm with three baseline methods, i.e., MARTO, MAMCC, and MAGCS. As shown in Fig. \ref{The mutual channel interference}, the proposed MAML-MADFRL algorithm has lower channel interference because it can better adapt to the dynamic network environment such as stochastic task arrivals and time-varying LEO locations. Furthermore, the proposed MADFRL algorithm can help each user access the optimal channel to achieve the computation offloading, which is beneficial to reducing the mutual channel interference and improving the processed number of task bits. As MARTO method only randomly processes the task either users or remote LEO units, it cannot find the optimal task scheduling scheme. Moreover, MAMCC only equally allocates the computational resources for each agent and MAGCS cannot find the optimal channel access, which causes higher channel mutual interference.

\begin{figure*}[!t]
\subfigure[The network throughput versus transmission bandwidth.]{
\begin{minipage}[t]{0.5\linewidth}
\centering{}
\includegraphics[width=7cm,height=4cm]{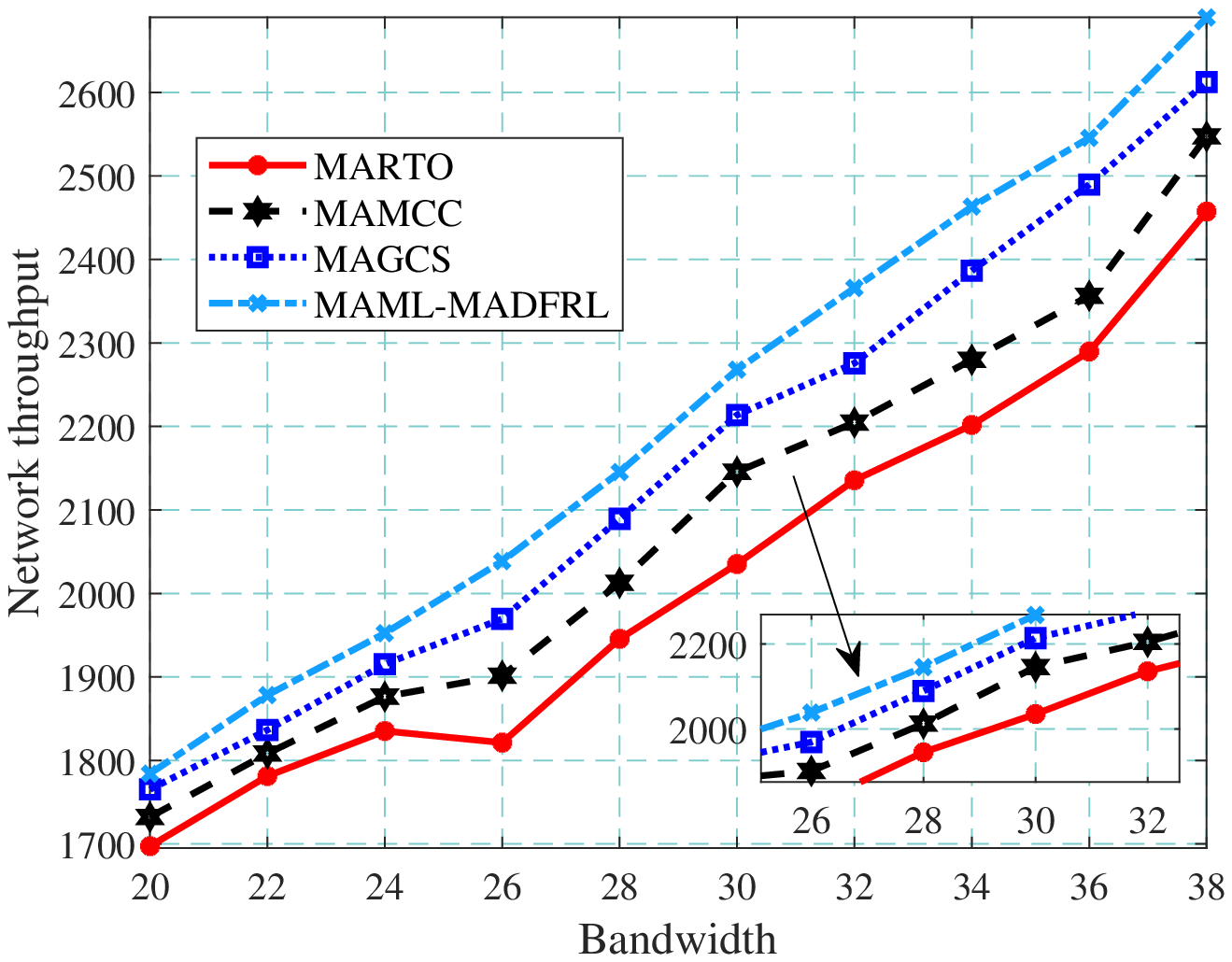}
\label{The network throughput versus transmission bandwidth}
\end{minipage}%
}%
\subfigure[The cloud servers' profits versus cloud servers' price.]{
\begin{minipage}[t]{0.5\linewidth}
\centering{}
\includegraphics[width=7cm,height=4cm]{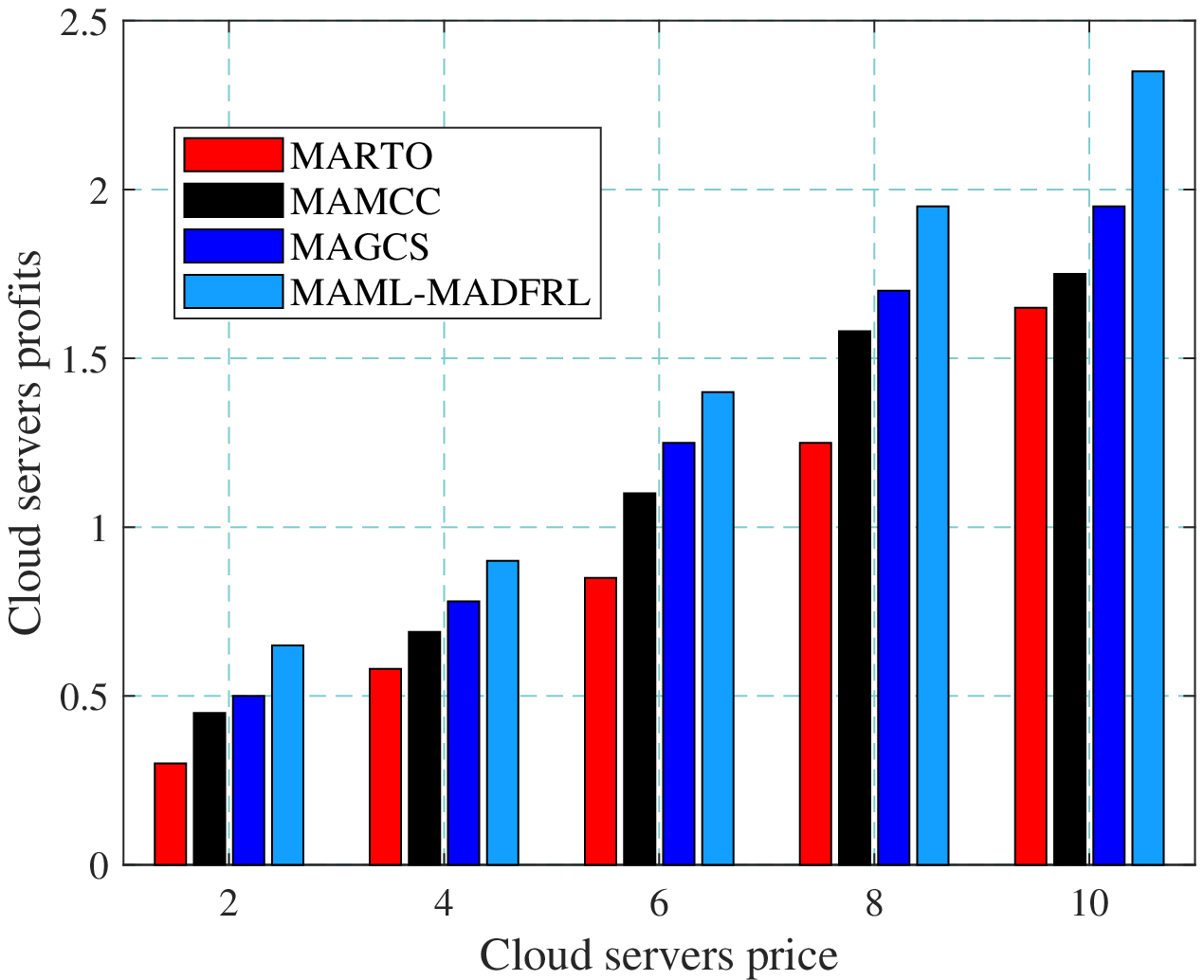}
\label{The cloud servers profits versus cloud server price}
\end{minipage}%
}%
\centering{}
\caption{The network throughput and cloud servers profits versus bandwidth and price.}
\end{figure*}

As shown in Fig. \ref{The network throughput versus transmission bandwidth}, we explore the impact of transmission bandwidth on network throughput. As transmission bandwidth increases, the average network throughput for all users gradually increases. Moreover, as the proposed MAML-MADFRL algorithm can better adapt to the stochastic task arrivals, time-varying LEO locations and adjust the cloud server price dynamically, this can obtain a larger network throughput compared with other three benchmarks. However, the channel bandwidth is limited in practical scenarios. Hence, we deploy the proposed MAML-MADFRL learning framework in practical SGIN systems to save bandwidth while guaranteeing the QoS, because meta learning can accelerate the training process from small batches of samples and the federated aggregation mechanism can help each user adjust the actor network parameters.
\ \ As shown in Fig. \ref{The cloud servers profits versus cloud server price}, we explore the impact of cloud servers' price on cloud servers' profits. Specifically, the proposed MAML-MADFRL learning framework has higher cloud servers profits because it can not only adapt to the dynamic network environment, but also adjust the LEO cloud pricing under different CPU cycle frequencies. Meanwhile, the proposed MAML-MADFRL learning framework can better solve the two-stage Stackelberg game model, which guarantees the cloud servers' profits and users' network throughput.
\vspace{0.02in}

\subsubsection{Privacy Protection Overhead}
\begin{figure*}[!t]
\subfigure[The privacy overhead versus model transmission factor.]{
\begin{minipage}[t]{0.5\linewidth}
\centering{}
\includegraphics[width=7cm,height=4cm]{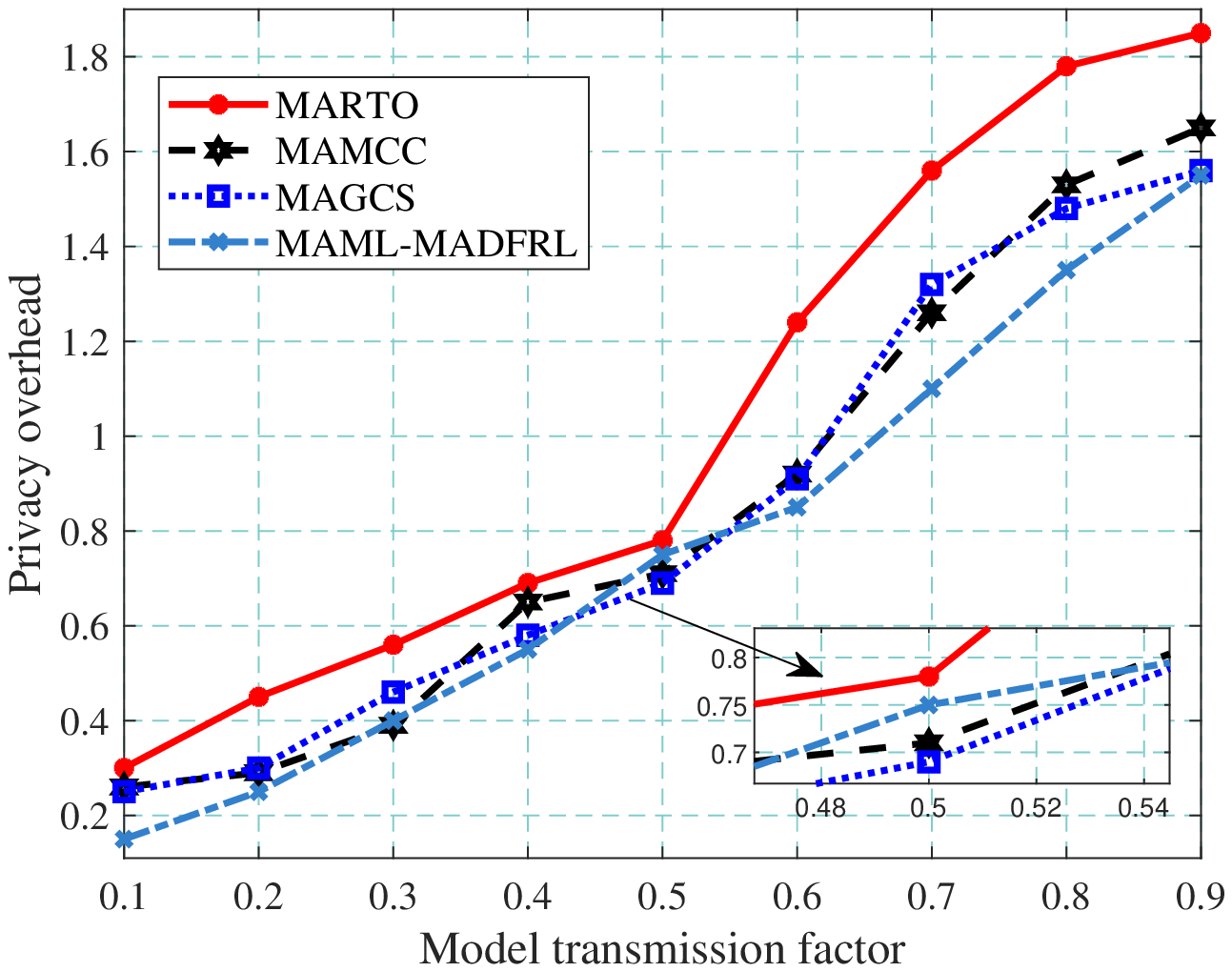}
\label{The privacy overhead versus model transmission factor}
\end{minipage}%
}%
\subfigure[The privacy overhead versus MBS computational capability.]{
\begin{minipage}[t]{0.5\linewidth}
\centering{}
\includegraphics[width=7cm,height=4cm]{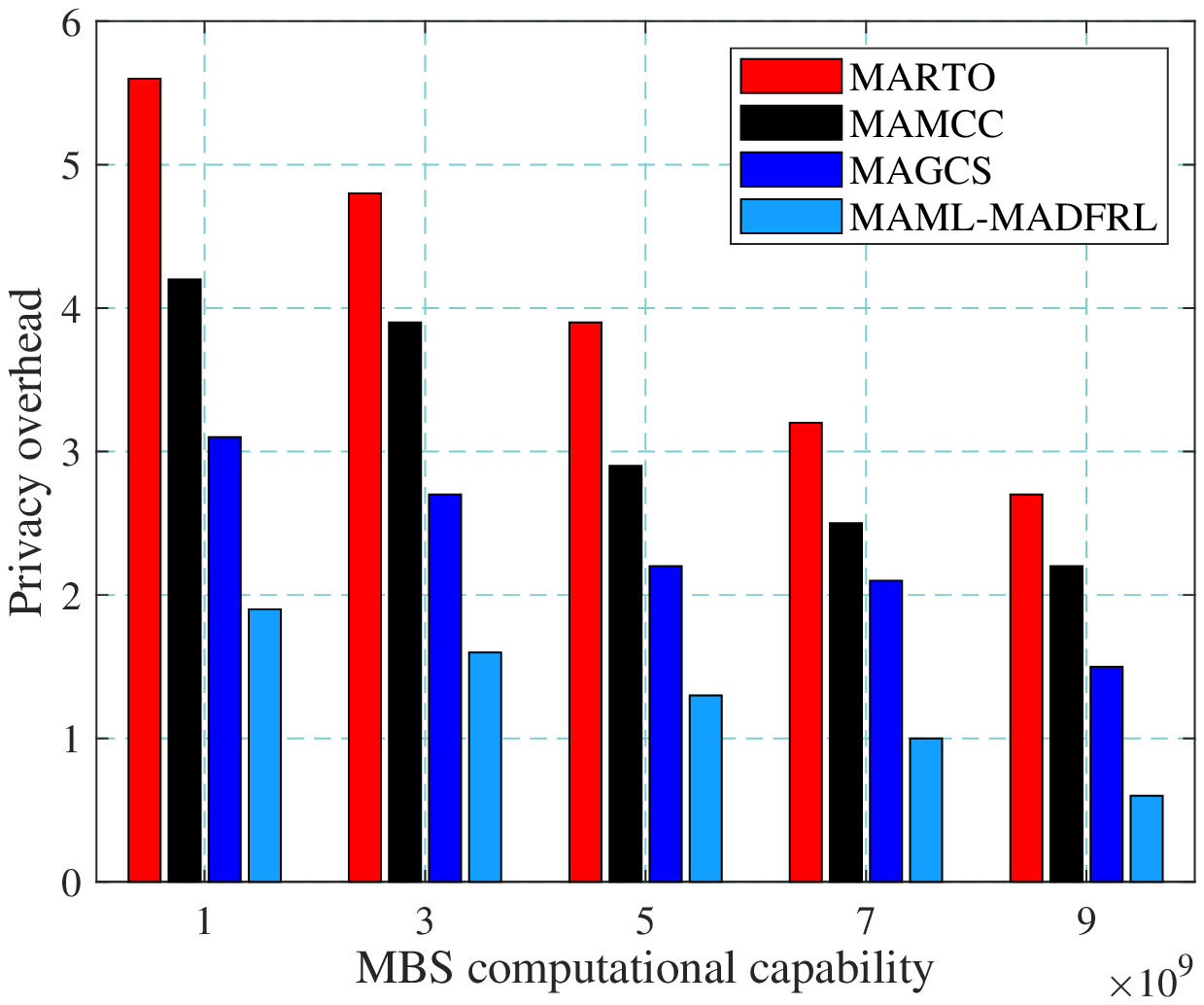}
\label{The privacy overhead versus MBS computational capability}
\end{minipage}%
}%
\centering{}
\caption{The privacy overhead versus model transmission factor and computational capability.}
\setlength{\belowdisplayskip}{0.1cm}
\end{figure*}

In this sequel, we explore the impact of model transmission factor on privacy protection overhead. As shown in Fig. \ref{The privacy overhead versus model transmission factor}, when the model transmission factor increases, the privacy overhead gradually increases. However, the proposed MAML-MADFRL algorithm has lower privacy overhead because the blockchain-aided federated aggregation and parameters issuing mechanism can help each user identify the network attack and choose the optimal block. Moreover, when the model transmission factor is greater than 0.5, the MARTO algorithm is hard to stabilize the privacy overhead because of random task scheduling. Hence, the proposed MAML-MADFRL learning framework can better adapt to model transmission factors and reduce privacy protection overhead.

Next, we explore the impact of MBS computational capability on the privacy overhead. As shown in Fig. \ref{The privacy overhead versus MBS computational capability}, when the computational capability of MBS gradually increases, the privacy overhead decreases, because the MBS with high computational capability accelerates network parameters aggregation. Moreover, the proposed MAML-MADFRL algorithm has lower privacy overhead compared to other baseline methods, because the MAML-MADFRL can not only speed up the parameters aggregation, but also allocate the optimal block for each user, which further reduce the privacy protection overhead.

\vspace{-0.1em}
\section{Conclusions}
\ \ In this paper, we consider a blockchain-aided two-stage Stackelberg game model in SGIDTNs scenarios to maximize the network throughput and cloud servers' profits in terms of minimum privacy protection overhead, stochastic task arrival, time-varying LEO locations and variables coupling for long-term task queue constraints and short-term computation offloading. Next, we propose a Lyapunov stability theory-based MAML-MADFRL learning framework to process the task scheduling, reduce the channel interference, optimize the CPU cycle frequency, and allocate the block size, which further achieve the optimum cloud servers profits via optimizing the cloud servers' prices. Moreover, we analyse the corresponding blockchain verification mechanism, the computational complexity of the proposed MAML-MADFRL algorithm, and algorithm performance bounds as well as task queue convergence. Finally, massive simulation results show that the proposed MAML-MADFRL learning framework achieves higher network throughput, cloud servers' profits, and lower privacy verification overhead compared to the MARTO, MAMCC and MAGCS benchmarks.

\bibliographystyle{IEEEtran}
\bibliography{reference5}

\end{document}